# Textures in Polygonal Arrangements of Square Nanoparticles in Nematic Liquid Crystals Matrices

[a]Paul M. Phillips, [b]N. Mei, [a,c]Ezequiel R. Soulé, [b]Linda Reven and [a*]Alejandro D. Rey.

[a]Department of Chemical Engineering, McGill University, 3610 University Street, Montreal, Quebec H3A2B2, Canada

[b]Center for Self-Assembled Chemical Structure (CSACS), Chemistry Department, McGill University, 801 Sherbrooke St. West, Montreal, Quebec H3A 2K6, Canada

[c]Institute of Materials Science and Technology (INTEMA), University of Mar del Plata and National Research Council (CONICET), J. B. Justo 4302, 7600 Mar del Plata, Argentina.

email: alejandro.rey@mcgill.ca

**Abstract**

A systematic analysis of defect textures in facetted nanoparticles with polygonal configurations embedded in a nematic matrix is performed using the Landau-de Gennes model, homeotropic strong anchoring in a square domain with uniform alignment in the outer boundaries. Defect and textures are analyzed as functions of temperature T, polygon size R, and polygon number N. For nematic nanocomposites, the texture satisfies a defect charge balance equation between bulk and surface (particle corner) charges. Upon decreasing the temperature, the central bulk defects split and together with other -1/2 bulk defects, are absorbed by the nanoparticle's corners. Increasing the lattice size decreases confinement and eliminates bulk defects. Increasing the polygon number increases the central defect charge at high temperature and the number of surface defects at lower temperatures. The excess energy per particle is lower in even than in odd polygons, and it is minimized for a square particle arrangement. These discrete modeling results show for first time that even under strong anchoring, defects are attached to particles as corner defects, leaving behind a low energy homogeneous orientation field that favors nanoparticle ordering in nematic matrices. These new insights are consistent with recent thermodynamic approaches to nematic nanocomposites that predict the existence of novel nematic/crystal phases and can be used to design nanocomposites with orientational and positional order.

## I. Introduction

Nanotechnology based on liquid crystals includes templating for nanomaterial fabrication,[1,2] nanoparticle (NP) assembly with engineered order,[3] nanowire actuation,[4] morphology control in nano and microemulsions,[5] controlled plasmonic response using an arrayed gold nanodot-liquid crystalline composite for better color tuning,[6] sensors based on interactions between mesogens and biomacromolecules and viruses,[7] and flexoelectric actuators inspired by outer hair cells.[8] Novel applications of LC based nanocomposites strongly depend on the ability to control aggregation and spatial distribution of the particles in the matrix.[9-11] Liquid crystals (LC) have received much attention as dispersing medium for colloidal particles and NPs as a flexible method for generating and controlling self assembly into complex structures.[12-20] The diverse functionalities of thermotropic liquid crystals emerge from their orientational ordering at the nano and micron scale when observed within a temperature range intermediate between the solid and liquid phase.[8] Liquid crystals being anisotropic soft matter exhibit useful optical, electrical and mechanical properties[21] and their orientation can be influenced by external fields [22-24], and contact with structured surfaces and interfaces.[25]

A key feature of gold NP/LC composites and also LC-based sensors for viruses is the interaction of the mesogens with the particle surface, such that the molecular orientation in the vicinity of the particle is affected, producing a long range elastic response. The distortions of the director induced by spherical particles, and the elastic interactions between two or more spherical particles have been widely studied in the past.[26-31] These interactions can lead to phase separation, or to the formation of different self-assembled structures of particles. One way of controlling the effect of the surface on the mesogen configuration is through the chemical functionalization of the particle[32]. In the case of faceted NPs, in addition to any chemical effect, the geometric discontinuity propagates through the LC, giving rise to orientation distortions that have to be accommodated either through surface and bulk defect nucleation, or through NP/NP defect links. It has been shown experimentally[17,18] that the interactions of nanoparticles in LC can be tailored by controlling the nanoparticle's shape, and different types of self-assembled structures can be obtained. Based on predictions for single particles and pairs of particles,[33-35] it is expected that for nano scale faceted particles, neighbouring particles are linked through defect lines at smaller distances, while at larger distances the isolated particles absorb the geometric frustration in the form of surface corner defects, as opposed to bulk defects in colloidal particles. As the topological charge (or amount of orientation winding) of a surface defect $C_s$ on a faceted particle is just the ratio of the misorientation angle between two adjacent faces and $2\pi$, the defect an edge (3D) or a corner (2D) can absorb or emit is $\pm C_s$. For example for a square particle, where the relative angle between to adjacent phase is $\pi/2$, the surface defect charge associated with a corner is $C_s = \pm(\pi/2)/2\pi = \pm\pi/4$, where the sign depends on the director rotation when encircling the defect in a counterclocwise direction. Based on these observations, it is predicted that faceted particles have active sites through they can interact with other neigbouring particles which plays a role in the local structure of NP/LCs.

A second key feature of NP/LC composites and also of colloidal filled nematics, such as carbon fibers embedded in carbonaceous mesophases, is the presence of polygonal geometries in the particle's positional arrangements. It has been suggested that LCs may provide new routes to the formation of ordered arrays of NPs due to the presence of reversible elastic forces[31]. In addition, it has been shown that spherical as well as polygonal colloidal inclusions in a nematic liquid crystal can self-assemble forming complex one- and two-dimensional structures[19,20]. The thermodynamic solution model[36,37] of NP/LC composites (which consider spherical NP as a component of a continuous phase), based on two non-conserved order parameters for mesogen orientation (quadrupolar tensor order parameter $\mathbf{Q}$) and NP positional order (scalar order parameter $\sigma$) and a conserved order parameter for NP concentration ($\varphi$) shows that different types of phases can exist, depending on temperature-concentration (T, $\varphi$) conditions: isotropic (both components are disordered, $\mathbf{Q}=\mathbf{0}, \sigma=0$), nematic (the LC molecules show orientational order and NPs are randomly dispersed, $\mathbf{Q}\neq\mathbf{0}, \sigma=0$), crystal (colloidal crystal where the NPs exhibit positional order and the mesogens are randomly oriented, $\mathbf{Q}=\mathbf{0}, \sigma\neq 0$), and nematic crystal (the mesogenic molecules show orientational order and the NPs exhibit positional order, $\mathbf{Q}\neq\mathbf{0}, \sigma\neq 0$).

Figure 1a shows a generic phase diagram acording to the thermodynamic solution model, and figure 1b shows a schematic of nematic and nematic-crystal phases. This models treats this phase in a very generic and simplified way, but it is expected that, in principle, the textures and defect structure (and consequently the distortion energy), can be strongly affected by the geometry of the partcile arrangement. Hence predicting orientation textures and defect formation in crystal particle lattices is of importance in order to characterize and understand in detail the structure and interactions in this phase.

For micron and submicron circular fiber with polygonal arrangements and embedded in a nematic matrix, the defect charge C inside the particle arrangement follows Zimmer's texture rule:[38] $C=-(N-2)/2$, where N is the polygon number; for example for N=3, the disclination has strength -1/2, and for N=4, the charge is -1; we note that the minus sign in $C=-(N-2)/2$ corresponds to homeotropic anchoring conditions. Furthermore, experiments and theory[38] demonstrate that for colloidal particles there is an even and odd effect, such that for odd polygons (N=3,5) the defect cores are singular and of C=-1/2 and for even polygons (N=4,6) the defect cores are non-singular (escaped core). The singular C=-1/2 defects in odd polygons are always observed because higher order disclinations (-3/2, -5/2,…) split into -1/2 disclinations due to the lower elastic energy since the energy scales with the defect charge square: $(C)^2$. For smaller submicron particles under stronger confinement, the non-singular escape core disclinations found in even polygons become singular core.[38] Texture rules for NP/LCs , as developed in the present paper, will provide a better understanding on the structure and stability when both orientation and facetted particle polygonal arrangements.

The purpose of this work is to explore, through theory and numerical computations based on the standard Landau-de Gennes LC model, the orientation textures in 2D that arise for NP polygonal arrangements in a nematic LC at different temperatures and different NP densities. Only the idealized case of particle arrangements with perfect order is considered. Strong anchoring is assumed at the NP/LC interface; strong anchoring is expected to apply to mesogen-decorated NPs, where the mesogen surface density, molecular conformation and length are designed to promote homeotropic anchoring. The present two-phase (mesogen/NP) discrete model approach complements the spatially homogeneous thermodynamic solution model[36] since the former takes into account elastic interactions explicitly while the later incorporates them implicitly. In future work the thermodynamic solution model will be extended to facetted particles.

(a)
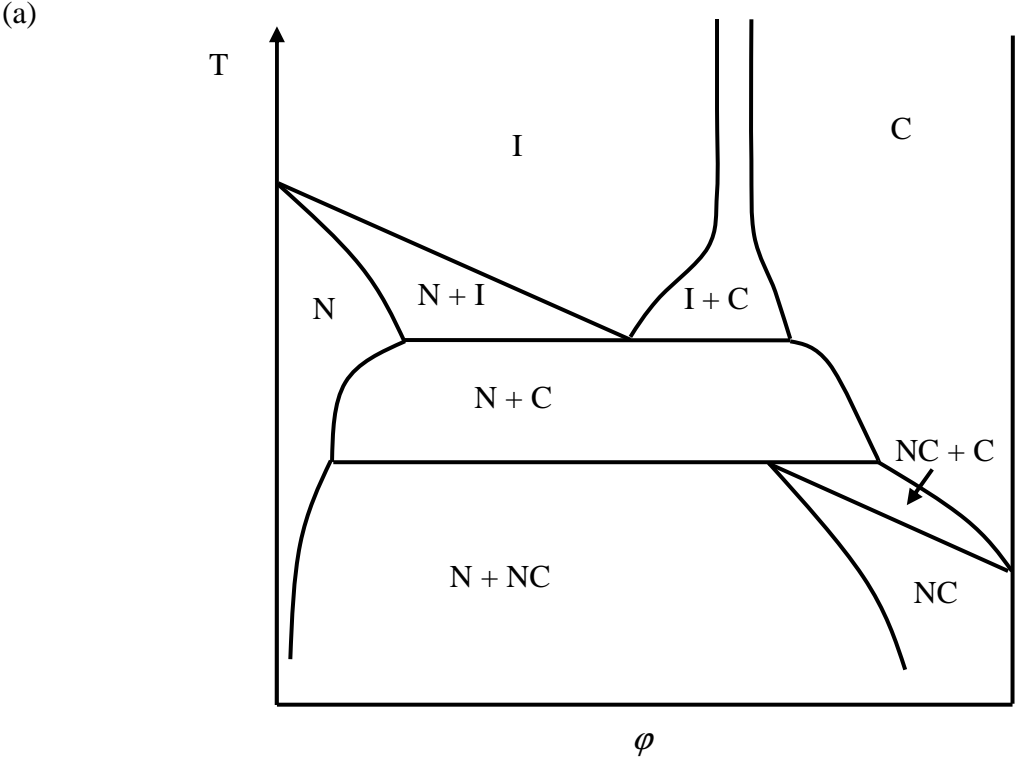

(b)
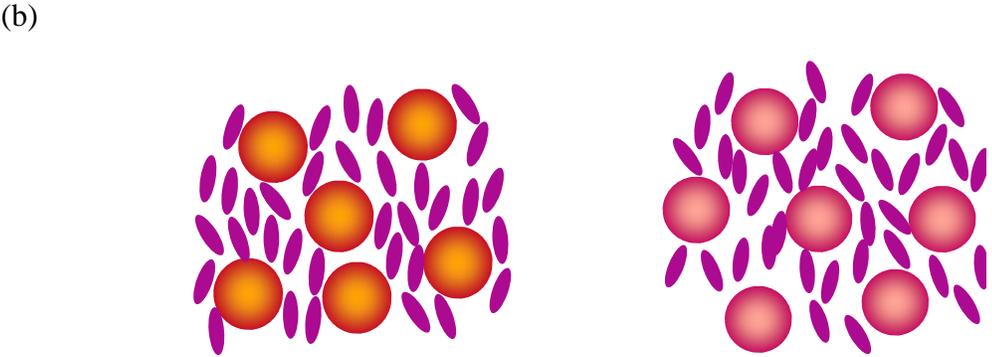

**Figure 1**. (a) Generic phase diagram of NP-LC dispersion, according to the thermodynamic solution model, in terms of temperature T and concentration of NPs $\varphi$. The different phases are: Isotropic (I), Nematic (N), Crystal (C), and Nematic-Crystal (NC). Detailed calculations of these phase diagrams and its dependence on physical and geometrical parameters can be found in refs. 26-27 (b) Schematic of a nematic nanocomposite formed by calamitic mesogens and spherical nanoparticles, showing the nematic phase (left), where the mesogens exhibit nematic ordering and the nanoparticles have random positional order, and the nematic-crystal phase (right): The mesogens exhibit nematic ordering and the nanoparticles have positional order.

**II. Theory and Simulation Methods**

The kinetics of texturing in NP/LC is studied using a square computational domain containing N square particles whose vertical and horizontal sides are aligned with the outer square domain (see Figure 2). Particle polygonal arrangements are defined in terms of the number N of square particles of constant size and include triangles (N=3), squares (N=4), pentagons (N=5), and hexagons (N=6). The textures are described by the symmetric traceless quadrupolar **Q**-tensor, parametrized in terms of orthonormal directors {**n,m,l**} and uniaxial and biaxial order parameters {S,P}: $\mathbf{Q} = \mathbf{Q}^T = S(\mathbf{nn} - \mathbf{I}/3) + P(\mathbf{mm} - \mathbf{ll})/3$; **n** is the main director or optic axis. Note that, even when a 2D computational domain is used, **Q** is a 3D tensor order parameter; in this way biaxiality (very important for the description of defects) can be accounted for. Strictly speaking, our model represents a 3D system wich is infinite and uniform in the direction perpendicular to the simulation plane. This type of modelling can still capture the main feature of real, finite 3D systems.

The evolution of **Q** is given by the torque balance equation (nematodynamics) in the absence of flow:[39]

$$\beta\left(\frac{\partial \mathbf{Q}}{\partial t}\right)_\mathbf{x} = -\left(\frac{\partial (f_h + f_g)}{\partial \mathbf{Q}}\right)^{[s]} + \left(\nabla \cdot \left(\frac{\partial f_g}{\partial \nabla \mathbf{Q}}\right)\right)^{[s]} \quad (1)$$

where β is the rotational viscosity coefficient, **Q** is the tensor order parameter that describes the ordering of the liquid crystal material,[15] t is time, [s] identifies that the matrices are symmetric and traceless, and $f_h$ and $f_g$ are, respectively, the homogeneous and gradient contributions to the free energy density of the system.[21] The free energy density of the isotropic phase, $f_{is}$, is used as the reference ($f_{is}$=0 J/m$^3$) and therefore does not appear in Eq.(1). Based on the classical Landau - de Gennes continuum theory for nematic liquid crystals,[21,40] the total free energy density f of the system can be expressed in terms of the tensor order parameter **Q** and its gradients:

$$f = f_h + f_g \tag{2a}$$

$$f_h(\mathbf{Q}) = \frac{a_o(T-T^*)}{2}\mathbf{Q}:\mathbf{Q} - \frac{b}{3}\mathbf{Q}:(\mathbf{Q}\cdot\mathbf{Q}) + \frac{c}{4}(\mathbf{Q}:\mathbf{Q})^2 \tag{2b}$$

$$f_g(\nabla\mathbf{Q}) = \frac{L_1}{2}\nabla\mathbf{Q}\vdots(\nabla\mathbf{Q})^T + \frac{L_2}{2}(\nabla\cdot\mathbf{Q})\cdot(\nabla\cdot\mathbf{Q}) \tag{2c}$$

where T is the quench temperature, $T^*$ is the clearing point temperature i.e. the spinodal temperature below which the isotropic state is unstable,[41] and the experimentally measured material parameters are $a_o$, b, and c (Landau coefficients) and $L_1$, $L_2$ (Landau elastic constants). Here for simplicity we adopt the equal bend/splay approximation. The material parameters used here (given in ref. 33) are for the liquid crystal 5CB (4-n-4'-pentyl-4-cyanobiphenyl) undergoing an isotropic-to-nematic phase transition at a temperature T below the supercooling T of the nematic state. 5CB is a well-studied nematic liquid crystal often used to investigate and create new nanotechnologies. For 5CB the phase stability ranges are: $T< T^*=307.2$ K, the isotropic (nematic) state is unstable (stable), for $T^*<T< T_c=307.4$ K, the isotropic (nematic) state is metastable(stable), for $T_c <T< T_u=307.47$ K, the isotropic (nematic) state is stable (metastable), for $T> T_u$ the isotropic (nematic) state is stable (unstable). In this paper we studied the range $300K < T < 307.4K$; for brevity below we omit K when indicating T. The external length scales are the particle box H, the NP size is L, the NP-NP separation distance is R, and the internal (coherence) length scale is $\xi = \sqrt{L_1/a_o} \approx 10$nm. The values H = 400nm and L = 40nm were adopted, while different values of R were analyzed. Following the procedure defined in ref. 33, we re-scale the equations, initial conditions, boundary conditions, and computational domain using the box size H and a characteristic time $\tau = \beta/a_0T^* = 2.10^{-9}$s, and solve eqn.(1) for $\mathbf{Q}(x,y,t)$ as a function of dimensionless coordinates (x,y) and dimensionless time t. The dimensionless computational domain consists of an outer unit square, $(0, 1)\times(0, 1)$ containing N square particles in a polygonal arrangements. The side of the polygon R is the interparticle dimensionless distance which is a variable parameter that is used to characterize the effect of packing on textures. For each polygon N, the particles are arranged for several R, and the centre point of the particle distribution is placed on the centre point of the computational domain. Fig. 2 displays a schematic of the triangular distribution.

The initial conditions used to solve eqn. (1) are the isotropic state. The boundary conditions on the outer unit square describe an equilibrium uniaxial vertically aligned nematic:

$$\mathbf{Q} = S_{eq} \left( \boldsymbol{\delta}_y \boldsymbol{\delta}_y - \mathbf{I}/3 \right) \tag{3a}$$

$$S_{eq} = b \left[ 1 + \sqrt{1 - 24 a_o c (T - T^*)/b^2} \right] / 4c \tag{3b}$$

where $\boldsymbol{\delta}_y$ is the unit vector in the vertical (y) direction and $S_{eq}$ is the equilibrium scalar order parameter. At the inner boundaries (sides of all square particles):

$$\mathbf{Q} = S_{eq} \left( \mathbf{kk} - \mathbf{I}/3 \right) \tag{3c}$$

where $\mathbf{k}$ is the unit normal.

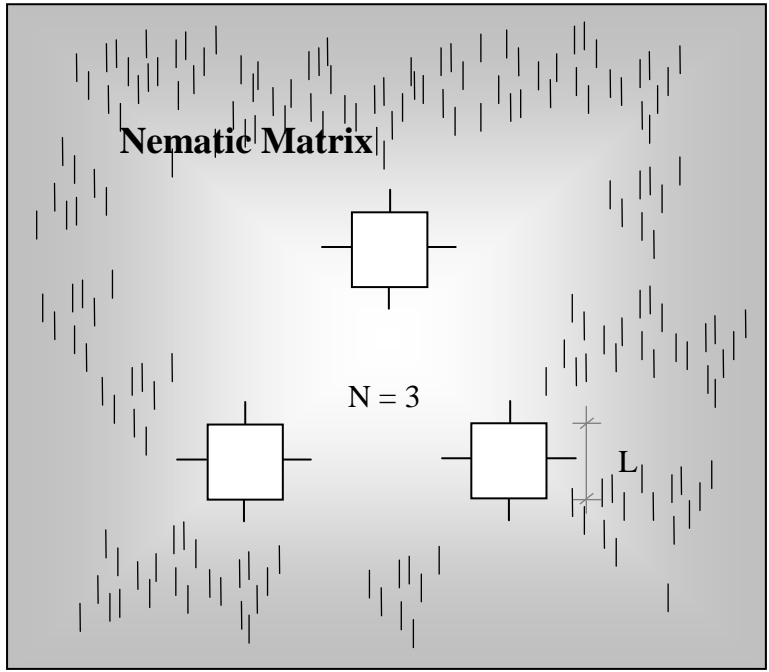

**Figure 2.** Schematic of the triangular (N=3) arrangement of nanoparticles (NP) within the computational domain of size H. In a given simulation the particles are arranged such that the interparticle distance between each neighbouring particles are kept at a constant R. The centre point of the particles arrangements lies on the centre point of the computational domain. The director **n** is normal to the NPs and is kept vertical ($\mathbf{n} = \boldsymbol{\delta}_y$) on the entire outer boundary.

**III. Results and Discussion.**

The number, strength and sign of all defects predicted by eqn.(1) for N=3,4,5,6 polygons as a function of quenching temperature T and dimensionless separation distance R are shown in Table I. At these small scales we note that there are no non-singular (escape core) defects. The first column defines the polygonal arrangement N, the second the temperature T, the third the interparticle distance R, the fourth the number (charge) of bulk defects, the fifth the number (charge) of surface defects, and the sixth column is the number (charge) associated with the particles. The defect charge equation that shows the balance between bulk and surface defects in a unit polygonal cell, satisfied at any (T,R,N) is:

$$\underbrace{N_b \times C_b}_{\text{total bulk charge}} + \underbrace{4 \times N \times C_s}_{\text{total surface charge}} \equiv \underbrace{N_b \times C_b}_{\text{total bulk charge}} + \underbrace{N \times C_p}_{\text{total particle charge}} = 0 \qquad (4)$$

where $\{C_i; i = b, s, p\}$ denotes bulk, surface, and particle charge, and $N_b$ denotes bulk defects. In the first term, the defects inside and outside the polygon are included. The particle charge is the sum of the surface defect charges on each particle and can be zero or positive; the left and right euqations (4) are equivalent. For example, Table I shows that for T=307.42, R=0.2, N=6, the bulk, surface and particle charges are:

$$N_b \times C_b = 8(-1/2) + 1(-2) = -6, \quad 4 \times N \times C_s = 4 \times 6 \times (+1/4) = +6, \quad N \times C_p = 6(+1)$$

The total charge balance balance euqation (4) then is -6+6=0. The two key texture processes, observed by changing the temperature T, polygon number N, and nanoparticle distance R, that lower the total homogeneous and gradient energy are:

*(a) Bulk defect splitting:* the observed defect splitting (indicated by a right arrow) cases include:

$$(-2) \to 4 \times (-1/2), \ (-3/2) \to 3 \times (-1/2), \ (-1) \to 2 \times (-1/2)$$

According to the values of (T,N,R) in this bulk defect splitting mode it is found that (i) a bulk defect of strength -2 splits into 4 bulk defects of strength -1/2, (ii) a -3/2 splits into three (-1/2), and (iii) a

-1 splits into two -1/2 . For example for (T,R,N)=(307.42,0.2,5) a -3/2 bulk defect splits into 3 (-1/2) defects by increasing R up to 0.3.

*(b) Bulk defect absorption by nanoparticle corners:* In this mode a negatively charged bulk defect migrates to the corner of a nanoparticle, reacts with a positively charged corner defect and leaves behind a negatively charged corner defect. The observed case involves:

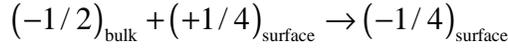

$$(-1/2)_{bulk} + (+1/4)_{surface} \rightarrow (-1/4)_{surface}$$

where a -1/2 bulk defect is absorbed by a corner whose charge is then changed from +1/4 to -1/4. For example for (T,R,N)=(307.42,0.4,5) ten -1/2 bulk defect are absorbed into corners by decreasing T to 307.3.

The sensitivity of defect splitting and defect absorption to the three parameters T, N, and R, shown in Table I, are as follows.

*(1) Temperature T effects:* Results for T=300 are essentially identical to T=307, indicating athermal behaviour at sufficiently low T. Bulk defects of charges -1/2,-1,-3/2,-2 are found at T=307.42. Zimmer's rule (bulk defect charge as a function of polygon number : $C_b$=-(N-2/2); (see above and ref. 38) is obeyed at the higher temperature T=307.42 only. For example, for T=307.42, R=0.2, N=6 there is one bulk defect inside the polygon, with $C_b$=-2 (Zimmer's rule). At the higher T=307.42, the defect charge balance eqn.(4) is:

$$N_{bo} \times \left(-\frac{1}{2}\right) - \frac{(N-2)}{2} + 4 \times N \times \frac{1}{4} = 1 + \frac{N - N_{bo}}{2} = 0 \tag{5}$$

where $N_{bo} = N + 2$ denotes the number of bulk -1/2 defects outside the polygon. For T<307.3 all the defects are at the surface and of the type +1/4 or -1/4 and the defect charge eqn.(4) is:

$$0 + 2 \times N \times \frac{1}{2} + 2 \times N \times \left(-\frac{1}{2}\right) = 0 \tag{6}$$

*(2) Polygon order N effects:* N affects the number of corners and particles which enter in the defect charge balance eqn. (4). According to Zimmer's rule, the central defect charge is –(N-2)/2, and hence increasing N increases this charge, as observed at higher temperature.

*(3) Polygon side length R effects:* The confinement which controls splitting of bulk defects and absorption by corners is quantified by the polygon's circumscribing circle of radius $r_c$:

$$r_c = \frac{R}{2\tan(\pi/N)} \qquad (7)$$

Hence increasing R at fixed N, decreases confinement, promotes bulk defect splitting (more lower charge defects) and bulk defect absorption. A key finding is that in the nematic phase (except at the higher temperature T=307.42), regardless of R and N, the corners absorb bulk defects.

To emphasize the role of temperature T, polygon number N, and confinement R, we show in figure 3 representative orientation plots that highlight the essential features presented in Table I. For fixed N and R, decreasing temperature eliminates bulk defects. For fixed T and N, increasing R eliminates bulk defects. For fixed T and R, increasing N increases the central defect bulk charge (Zimmers effect) at high T and the number of surface defects at low T.

**Table 1.** Defect Structures. N: polygon number, T: temperature, R: polygon side length, X(Y): number of defects (sign and charge).

|  |  | R | Bulk | Surface | Particle Charge |
|---|---|---|---|---|---|
| N = 3 | T = 307.42K | 0.2 | 6(−½) | 12(+¼) | 3(+1) |
|  |  | 0.3 | 2(−½) 1(−1) | 10(+¼) 2(−¼) | 2(+1) 1(0) |
|  |  | 0.4 | 0 | 6(+¼) 6(−¼) | 3(0) |
|  | T = 307.3K | 0.2 | 0 | 6(+¼) 6(−¼) | 3(0) |
|  |  | 0.3 | 0 | 6(+¼) 6(−¼) | 3(0) |
|  |  | 0.4 | 0 | 6(+¼) 6(−¼) | 3(0) |
|  | T = 307.0K | 0.2 | 0 | 6(+¼) 6(−¼) | 3(0) |
|  |  | 0.3 | 0 | 6(+¼) 6(−¼) | 3(0) |
|  |  | 0.4 | 0 | 6(+¼) 6(−¼) | 3(0) |
| N = 4 | T = 307.42K | 0.2 | 6(−½) 1(−1) | 16(+¼) | 4(+1) |
|  |  | 0.3 | 8(−½) | 16(+¼) | 4(+1) |
|  |  | 0.4 | 8(−½) | 16(+¼) | 4(+1) |
|  | T = 307.3K | 0.2 | 0 | 8(+¼) 8(−¼) | 4(0) |
|  |  | 0.3 | 0 | 8(+¼) 8(−¼) | 4(0) |
|  |  | 0.4 | 0 | 8(+¼) 8(−¼) | 4(0) |
|  | T = 307.0K | 0.2 | 0 | 8(+¼) 8(−¼) | 4(0) |
|  |  | 0.3 | 0 | 8(+¼) 8(−¼) | 4(0) |
|  |  | 0.4 | 0 | 8(+¼) 8(−¼) | 4(0) |
| N = 5 | T = 307.42K | 0.2 | 7(−½) 1(−3/2) | 20(+¼) | 5(+1) |
|  |  | 0.3 | 10(−½) | 20(+¼) | 5(+1) |
|  |  | 0.4 | 10(−½) | 20(+¼) | 5(+1) |
|  | T = 307.3K | 0.2 | 0 | 10(+¼) 10(−¼) | 5(0) |
|  |  | 0.3 | 0 | 10(+¼) 10(−¼) | 5(0) |
|  |  | 0.4 | 0 | 10(+¼) 10(−¼) | 5(0) |
|  | T = 307.0K | 0.2 | 0 | 10(+¼) 10(−¼) | 5(0) |
|  |  | 0.3 | 0 | 10(+¼) 10(−¼) | 5(0) |
|  |  | 0.4 | 0 | 10(+¼) 10(−¼) | 5(0) |
| N = 6 | T = 307.42K | 0.2 | 8(−½) 1(−2) | 24(+¼) | 6(+1) |
|  |  | 0.3 | 8(−½) 4(−½) | 24(+¼) | 6(+1) |
|  |  | 0.4 | 0 | 12(+¼) 12(−¼) | 6(0) |
|  | T = 307.3K | 0.2 | 0 | 12(+¼) 12(−¼) | 6(0) |
|  |  | 0.3 | 0 | 12(+¼) 12(−¼) | 6(0) |
|  |  | 0.4 | 0 | 12(+¼) 12(−¼) | 6(0) |
|  | T = 307.0K | 0.2 | 0 | 12(+¼) 12(−¼) | 6(0) |
|  |  | 0.3 | 0 | 12(+¼) 12(−¼) | 6(0) |
|  |  | 0.4 | 0 | 12(+¼) 12(−¼) | 6(0) |

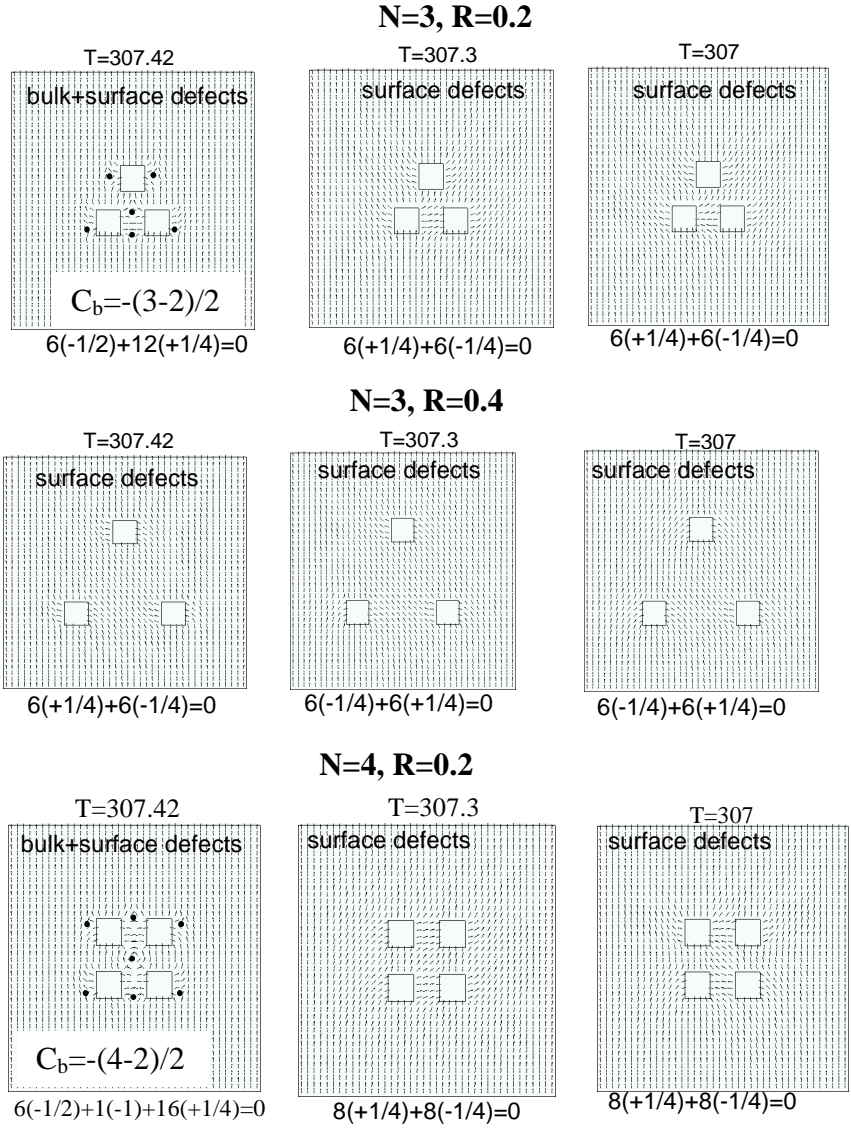

**Figure 3.** Representative computed director **n** visualization as a function of temperature T, polygon size R, and polygon number N. Realizations of the texture balance eqn. (4) are indicated below each frame. The Zimmer's bulk charge $C_b=-(N-2)/2$ is indicated in the lower left of the first and third row. Small black circles indicate the position of bulk defects.

In addition to defect and texture processes in LC nanocomposites, the second objective is to identify possible polygonal odd-even effects that may lower the elastic energy. When a particle is introduced in a nematic matrix, the director is distorted in the vicinity of the particle. If the particles are far away from each other, their corresponding distortion areas do not interact, but when they are close enough, the distortion areas overlap and interact to minimize the total distortion. For example, when four particles forming a square arrangement under small inter-particle separation the selected surface defect mode (T<307.42), contains a region of relatively homogeneous director orientation in the interior of the square (lower panel in figure 3). This elastic interaction is geometry-dependent and can be characterized by the excess free energy per particle:

$$F_e = \frac{\int_V (f_h + f_g) dV - \int_V f_{h,ref} dV}{N} \qquad (8)$$

where the integration is carried over the volume of the LC, and $f_{h,ref}$ is the free energy of an undistorted nematic LC. This quantity represents the increase of total free energy when a particle is added to a nematic LC and is essentially the insertion energy. Figure 4 shows $F_e$ as a function of N, for different distances R and temperatures T. In general, decreasing the separation between particles permits a higher interaction and $F_e$ thus is lowered. A non-monotonic dependence of $F_e$ on N can be observed in figure 4. Polygons with even number of particles seem to be energetically preferred (altough there are deviations due to the effect of the boundaries of the simulation domain for large values of N and R), and in particular, the square arrangement has the minimum energy.

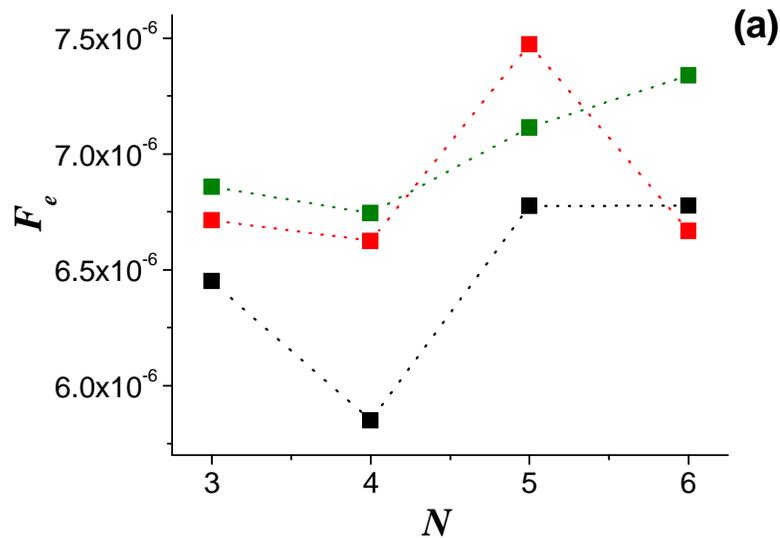

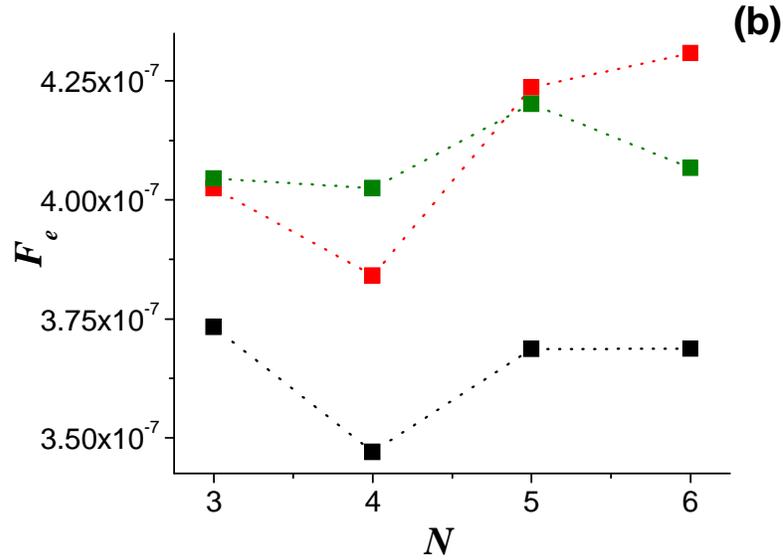

**Figure 4.** Excess free energy per particle, as a function of the number of particles. (a) T=300, (b) T=307.3. Black: R=0.2, red: R=0.3, green: R=0.4.

**IV. Conclusion**

In summary, a systematic analysis of defect textures in faceted particles with polygonal configurations embedded in a nematic matrix was performed using the Landau-de Gennes model, for different temperatures, polygon sizes and number of particles, in the ideal case of arrangements of particles with perfect order. Bulk defects were found to be stable at high temperature and small polygon size, while surface defects are favored at low temperatures and large polygon sizes. Increasing the polygon number increases the central defect $C_b=-(N-2)/2$ at high temperature and the number of surface defects at lower temperatures, according to Zimmer's texture rule. The excess energy per particle was calculated, and it was found to be geometry-dependent: it is lower in even than odd polygons, and it is minimized for a square arrangement. These discrete modeling results complements the continuous thermodynamic model that predicts the existence of nematic-crystal phases in nematic-NP composites, and shows that even under strong anchoring defects are attached to particles leaving behind a healed orientation field. These new insights represent an important step towards the rational design of novel nanocomposites with orientational and positional order.

**Acknowledgements**

This work is partially supported by a grant from FQNRT (Quebec). ADR gratefully acknowledges partially support by a grant from the Petroleum Research Fund of the American Chemical Society, and by the U.S. Office of Basic Energy Sciences, Department of Energy, grant DE-SC0001412.

**For Table of Contents only**

*central defect charge* C=-(N-2)/2

*N=3, 5*

*N=2, 4*

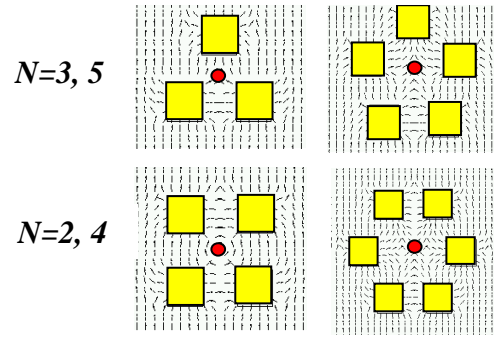